\documentclass[prd,a4paper,showpacs,nofootinbib]{revtex4}

\usepackage{epsfig}
\usepackage{amsmath}

\newcommand{\be}{\begin{equation}}
\newcommand{\ee}{\end{equation}}
\newcommand{\ba}{\begin{eqnarray}}
\newcommand{\ea}{\end{eqnarray}}

\newcommand{\simorder}{\raisebox{-4pt}{$\, \stackrel{\textstyle >}{\sim} \,$}}

\newcommand{\lf}{\left}
\newcommand{\rg}{\right}

\graphicspath{{plots/}}

\begin{document}

\title{\bf The saturation scale and its {\boldmath$x$}-dependence from 
{\boldmath$\Lambda$} polarization studies}

\author{Dani\"el Boer}
\email{D.Boer@few.vu.nl}
\affiliation{Department of Physics and Astronomy,
Vrije Universiteit Amsterdam, \\
De Boelelaan 1081, 1081 HV Amsterdam, The Netherlands}

\author{Andre Utermann\footnote{Present address: Institut f\"ur
    Theoretische Physik, Universit\"at Regensburg, D-93040 Regensburg, 
Germany}}
\email{Andre.Utermann@physik.uni-regensburg.de}
\affiliation{Department of Physics and Astronomy,
Vrije Universiteit Amsterdam, \\
De Boelelaan 1081, 1081 HV Amsterdam, The Netherlands}

\author{Erik Wessels}
\email{E.Wessels@few.vu.nl}
\affiliation{Department of Physics and Astronomy,
Vrije Universiteit Amsterdam, \\
De Boelelaan 1081, 1081 HV Amsterdam, The Netherlands}

\date{\today}

\begin{abstract}
  The transverse polarization of forward $\Lambda$ hyperons produced
  in high-energy $p\,$-$A$ collisions is expected to display an
  extremum at a transverse momentum around the saturation scale. This
  was first observed within the context of the McLerran-Venugopalan
  model which has an $x$-independent saturation scale. The extremum
  arises due to the $k_t$-odd nature of the polarization dependent
  fragmentation function, which probes approximately the derivative of
  the dipole scattering amplitude. The amplitude changes most strongly
  around the saturation scale, resulting in a peak in the
  polarization. We find that the observation also extends to the more
  realistic case in which the saturation scale $Q_s$ is $x$ dependent.
  Since a range of $x$ and therefore $Q_s$ values is probed at a given
  transverse momentum and rapidity, this result is {\it a priori} not
  expected. Moreover, the measurement of $\Lambda$ polarization over a
  range of $x_F$ values actually provides a direct probe of the $x$
  dependence of the saturation scale. This novel feature is
  demonstrated for typical LHC kinematics and for several
  phenomenological models of the dipole scattering amplitude.  We show
  that although the measurement will be challenging, it may be
  feasible at LHC.  The situation at RHIC is not
  favorable, because the peak will likely be at too low transverse
  momentum of the $\Lambda$ to be a trustworthy measure of the
  saturation scale.

\end{abstract}


\maketitle

\section{Introduction}
It is well-known that $\Lambda$ hyperons produced in collisions of
unpolarized hadrons are to a large degree polarized perpendicularly to
the production plane. Even though the origin of this phenomenon has
not been clarified fully yet, for sufficiently large transverse
momentum $p_t$ of the $\Lambda$, one expects that a parton description
must be applicable.  In Ref.~\cite{ABDM1} it was shown that the
available $\Lambda$ polarization data for $ p_t > 1$ GeV can be
described within a factorized approach by employing a polarization and
transverse momentum dependent fragmentation function. This function,
denoted by $\Delta^N D$, describes the fragmentation of an unpolarized
quark into a transversely polarized $\Lambda$ and is an odd function
of the transverse momentum $k_t$ of the quark w.r.t.\ the $\Lambda$
momentum \cite{Mulders:1995dh}. This $k_t$-odd nature implies that it
is essentially accompanied by the first derivative of the partonic
cross section w.r.t.\ $k_t$, unlike the unpolarized $\Lambda$
fragmentation function $D$, which is $k_t$-even.  This turns the
observed $\Lambda$ polarization into a useful tool, which in this
paper will be applied to the study of the $x$-dependence of the
saturation scale.

Once the polarization dependent fragmentation function is known from
data in the large $x$ or DGLAP region, polarization measurements in
other kinematic regions could point to changes in the underlying
physics. In Ref.\ \cite{Boer:2002ij} this was discussed specifically
for the saturation region, the region of small momentum fraction $x$
where the gluon density is very high and is expected to saturate
according to the nonlinear evolution equations of relevance in that
region. The saturation region is characterized by the so-called
saturation scale $Q_s$, i.e.\ the momentum scale at which saturation
effects become sizable. It was noted within the context of the
McLerran-Venugopalan model \cite{MV}, which has an $x$-independent
saturation scale $Q_s$, that the negative valued $\Lambda$
polarization displays a minimum at a transverse momentum approximately
equal to $Q_s$. Here, we want to investigate the more realistic case
in which the saturation scale is $x$ dependent
\cite{Mueller:1999wm,Mueller:2002zm,Triantafyllopoulos:2002nz}.  It
may be expected that, since now a range of $Q_s$-values is probed, the
minimum of the polarization is smeared out, and possibly not
recognizable anymore. However, in this paper we will demonstrate that
this is not the case. In fact, the pronounced minimum of the
polarization can even be used to probe the $x$ dependence of the
saturation scale. This makes the observable of potential interest for
collider experiments at RHIC, LHC and a future electron-ion collider,
the EIC.

Having discussed this promising use of $\Lambda$ polarization, let us
now address the possibilities of measuring it in the small-$x$
region. In high energy scattering the polarization of a spin-1/2 final
state hadron can usually only be measured through self-analyzing
parity violating decays. Exploiting this property it was demonstrated
already more than 30 years ago \cite{Lesnik:1975my,Bunce:1976yb} that
$\Lambda$ hyperons emerging from unpolarized $p\,$-$A$ collisions are
polarized perpendicularly to the production plane (cf.\ Ref.\
\cite{Panagiotou:1989sv} for an extensive review of data). In the
fixed target experiments performed at typical center of mass energies
$\sqrt{s} \sim 20$ GeV, the transverse momentum dependence of the
degree of polarization shows the characteristic feature that after a
linear rise up to $p_t \sim 1$ GeV, it stays remarkably constant up to
the highest measured values $p_t \sim 4$ GeV. This behavior was found
to be independent of the specific values of $\sqrt{s}$ and atomic
number $A$. For larger $p_t$ values one expects the asymmetry to fall
off as $1/p_t$, but this has not been observed yet. None of the
measurements performed thus far are in a kinematic region where the
target could be considered dense, i.e.\ in the saturation region.  In
Ref.\ \cite{Boer:2002ij} it was pointed out that the characteristic
flat behavior observed for $p_t$ values of a few GeV, would in that
case no longer be present, but rather an extremum should be visible,
located at $p_t \sim Q_s$. In other words, the observed plateau should
turn into a peak as saturation effects set in and $Q_s$ becomes a
  relevant scale. Since $Q_s$ grows with $1/x$
and $A$, one expects this to happen when $\sqrt{s}$ and/or $A$ are
increased significantly. In addition, it also helps to consider large
rapidities, i.e.\ forward $\Lambda$ production, in order to decrease
the $x$ values probed. At RHIC forward $\Lambda$'s with rapidities of
around 4 would begin to probe the small-$x$ region according to a
dipole scattering description \cite{djm2}. The possibilities to probe
small $x$-values are of course greater at LHC, where due to the
much higher $\sqrt{s}$ much less forward $\Lambda$'s are required.
For completeness, we recall that at RHIC $d\,$-$Au$ collisions have
been studied at energies of $\sim200$~GeV/$A$ in the nucleon-nucleon
center of mass frame. At LHC $p\,$-$Pb$ collisions will be performed
at $\sqrt{s_{NN}}=8.8$ TeV, but these do not take place in the
nucleon-nucleon center of mass frame which leads to a rapidity shift
from lab frame to center of rapidity frame of about half a unit.  In
principle also the $p\,$-$p$ collisions at LHC are of interest here,
due to the very large energy: $\sqrt{s}=14$ TeV.

Experimentally the measurement of forward $\Lambda$'s and their
polarization may be hampered by the often restricted particle
identification capabilities in the forward region. Two-thirds of the
time $\Lambda$'s decay into protons and negatively charged pions:
$\Lambda \to p\, \pi^-$. The angular distribution of the decay in the
$\Lambda$ rest frame is used to determine the polarization of the
$\Lambda$. Unfortunately, protons are usually hard to identify in the
forward region. In that case the only alternative may be to use that
the $\Lambda$'s decay one third of the time into neutrons and neutral
pions (and subsequently, two photons): $\Lambda \to n \, \pi^0 \to n
\, \gamma \gamma$.  Neutrons, $\pi^0$'s and photons have been
identified in the forward region at RHIC, hence this alternative may
be feasible \cite{Bland} and may in fact be the only way of measuring
$\Lambda$ polarization in the forward region at RHIC, LHC or EIC. We
will proceed with our investigation under the assumption that forward
$\Lambda$ polarization measurements will be possible in the future.

The outline of this paper is as follows. In section \ref{sect_plam} we
discuss the $\Lambda$ polarization asymmetry in terms of the relevant
polarization dependent fragmentation function and the dipole
scattering amplitude. This discussion repeats the essentials from
Ref.\ \cite{Boer:2002ij} in order to set the notation and to explain
why an extremum is expected at $p_t \propto Q_s$, but also it includes
the details of various phenomenological models for the dipole
scattering amplitude that were considered in the literature. As
mentioned, in Ref.\ \cite{Boer:2002ij} only the McLerran-Venugopalan
model was considered, but here we will focus on more recent models
that employ an $x$-dependent saturation scale. In section
\ref{sect_results} we discuss model results for the $\Lambda$
polarization observable for LHC kinematics mainly and point out the
generic qualitative features. Achieving realistic quantitative
predictions for the degree of $\Lambda$ polarization will not be our
aim, due to the large uncertainty in the polarization dependent
fragmentation function. Nevertheless, an estimate can be given of the
range of $x_F$ values required to observe the $x$ dependence of the
saturation scale, as the $p_t$ dependence of the $\Lambda$
polarization is found to be less model dependent than its absolute
value. Prospects for RHIC are also briefly discussed, but no results
will be shown. The reason for this is that at RHIC the peak is likely
situated at a $p_t$ below 1 GeV, where the considered
framework would not be appropriate. We end with conclusions.

\section{Transverse $\Lambda$ polarization description at small $x$}
\label{sect_plam}
According to Ref.\ \cite{Boer:2002ij}, the transverse polarization of
forward $\Lambda$'s produced in unpolarized $p\,$-$A$ collisions is
approximately given by
\begin{equation}
\mathcal{P}_\Lambda(p_t,x_F) = \frac
{\int_{x_F}^1 dx \, {x} \sum_q f_{q/p}(x,\mu_f^2) 
 \Delta^N D_{\Lambda^\uparrow/q}\left(\frac{x_F}{x},\mu_f^2\right) 
 \left[ N_F\left(\frac{x}{x_F}(p_t-k^0_t),x_2\right) - 
        N_F\left(\frac{x}{x_F}(p_t+k^0_t),x_2\right) \right]
}
{\int_{x_F}^1 dx \, {x} \left[\sum_q f_{q/p}(x,\mu_f^2) 
 D_{\Lambda/q}\left(\frac{x_F}{x},\mu_f^2\right) 
 N_F\left(\frac{x}{x_F}p_t,x_2\right)+ 
f_{g/p}(x,\mu_f^2)
D_{\Lambda/g}\left(\frac{x_F}{x},\mu_f^2\right) 
 N_A\left(\frac{x}{x_F}p_t,x_2\right)\right]
}\,,
\label{Plam}
\end{equation}
where $y_h$ and $p_t$ are the rapidity and the transverse momentum of
the produced $\Lambda$, $x_F=p_t/\sqrt{s}\,\exp[y_h]$, and $x$ and
$x_2=q_t^2/(x\,s)$ are respectively the momentum fractions of the
parton in the proton and the heavy ion (referred to as the
``target''). Note that three different values of $x_2$ enter in this
polarization expression, in conjunction with the three different
transverse momenta $q_t=x/x_F (p_t\pm k_t^0)$ and $q_t=x/x_F\, p_t$ of
the scattered parton.  The expression (\ref{Plam}) is based on the
asymmetry expression of Ref.\ \cite{ABDM1} combined with the dipole
picture description of the cross section of Ref.\ \cite{djm2}. In
the dipole formalism, $N_F$ describes the scattering of a quark off a
nucleus, while $N_A$ describes gluon-nucleus scattering. For details
of the derivation and justification of the approximations we refer to
Ref.\ \cite{Boer:2002ij}, where the dipole scattering amplitude $N_F$
is denoted by $C$.

The polarization dependent fragmentation function $\Delta^N
D_{\Lambda^\uparrow/q}$ is parameterized in terms of the unpolarized
one $D_{\Lambda/q}$ of Ref.\ \cite{deFlorian:1997zj},
\begin{equation}
\Delta^N D_{\Lambda^\uparrow/q}(z,\mu^2)
\equiv f_q^\Delta(z)\,D_{\Lambda/q}(z,\mu^2)\,,
\end{equation}
where
\begin{equation}
 f_{q}^\Delta=\tfrac{1}{2}\,N_{q} z^{c_{q}}(1-z)^{d_{q}}\,,\quad 
N_u=N_d=-28.13\,,\quad N_s=57.53\,,\quad c_q=11.64\,,\quad d_q=1.23\,,
\end{equation}
and the average transverse momentum is given by
\begin{equation}
 k_t^0(z)=0.66 z^{0.37} (1-z)^{0.5}\,{\rm GeV}\,.
\end{equation}
We emphasize that there is a large uncertainty in this
parameterization extracted from fixed target data \cite{ABDM1}, so
that the numerical results presented below should only be viewed as
qualitative, not as quantitative predictions. Future collider data
from LHC could be used to obtain a more trustworthy parameterization,
for instance through the $\Lambda$+jet observable recently pointed out
in Ref.\ \cite{Boer:2007nh}, which deals with $\Lambda$'s at
midrapidity where particle identification does not pose a problem.

Using the McLerran-Venugopalan (MV) model for the dipole scattering
amplitude it was shown in \cite{Boer:2002ij} that the
$p_t$ distribution of the transverse polarization displays a peak that
is directly related to the saturation scale $Q_s$. However, the MV
model does not incorporate evolution in $x$; $Q_s$ is constant. Here
we want to investigate whether $\mathcal{P}_\Lambda$ as a function of
$p_t$ still possesses an observable peak when described using a more
realistic dipole scattering amplitude including $x$-evolution. A
phenomenologically successful model with an $x$-dependent $Q_s$ is the
Golec-Biernat and W\"{u}sthoff (GBW) model \cite{GBW}, which was able
to describe small-$x$ DIS data\footnote{Note that the GBW model was
found to be inconsistent with newer, more accurate data and requires
some modification at larger $Q^2$, 
see for example Ref.\ \cite{Bartels:2002cj}.}.
Here we will focus on two different modifications of the GBW model
that were introduced to describe RHIC $d\,$-$Au$ data: the DHJ model
\cite{DHJ1,DHJ2} and the geometric scaling (GS) model of Ref.\
\cite{Boer:2007ug}. Both are of the form
\begin{equation}
N_F(q_t,x_2) \equiv \int d^2 r_t\: e^{i \vec{q}_t
    \cdot \vec{r}_t} \left[1-\exp\left(-\frac{1}{4}\left(\frac{4}{9}r_t^2
      Q_s^2(x_2)\right)^{\gamma(q_t,x_2)}\right) \right]~,
\label{NF_param}
\end{equation}
where $N_A$ is obtained from $N_F$ by replacing
$\tfrac{4}{9}r_t^2Q_s^2(x_2)$ by $r_t^2Q_s^2(x_2)$, and the saturation
scale is given by \cite{GBW}
\begin{equation}
Q_s(x_2) = 1\,\mathrm{GeV}\,
\left(\frac{x_0}{x_2}\right)^{\lambda/2},
\label{Qsx2}
\end{equation}
where the parameters $x_0 \simeq 3 \times 10^{-4}$ and $\lambda
\simeq 0.3$ were fitted to the small-$x$ DIS data. In the GBW model
the so-called anomalous dimension $\gamma$ is simply equal to 1. The
DHJ model incorporates expectations on the behavior of $\gamma$ from
BFKL/BK evolution \cite{DHJ1,DHJ2}, namely a logarithmic dependence on
$q_t^2/Q_s^2(x)$, and violations of geometric scaling that are
proportional to $1/y$ (at large $y$) 
\begin{equation} 
\gamma_{\rm DHJ}(q_t,x_2) = \gamma_s + (1-\gamma_s)\,
  \frac{|\log(q_t^2/Q_s^2(x_2))|}{\lambda
    y+d\sqrt{y}+|\log(q_t^2/Q_s^2(x_2))|}, 
\label{gammaDHJ}
\end{equation} 
where $\gamma_s=0.6275$, $y=\log 1/x_2$ is minus the rapidity of the
target parton. The saturation scale $Q_s(x)$ and the parameter
$\lambda$ are taken from the GBW model, as given in Eq.\ (\ref{Qsx2}),
and $d=1.2$ was fitted to data.  Here $Q_s$ includes the additional
factor $A^{1/3}$, where for large atomic numbers $A$ usually a
lower, effective number $A_{\rm eff}$ is used to account for impact
parameter dependence.

The parameterization of $\gamma_{\rm DHJ}$, which is based on
the one given in Ref.~\cite{Kharzeev:2004yx}, is well motivated by
expectations from small-$x$ evolution that are valid only for $q_t\ge
Q_s$ \cite{Mueller:2002zm,Iancu:2002tr}. 
 In contrast, the
continuation to the saturation region $q_t\le Q_s$ is rather
undetermined. In the case of hadron production at RHIC \cite{DHJ1,DHJ2}
this is not crucial since this region is hardly probed. But the
polarization observable discussed here is sensitive to $\gamma$ around
$Q_s$, not only to $q_t\ge Q_s$.  Hence, the continuation of $\gamma$
to the saturation region affects the polarization around the peak.

The DHJ model was found to describe well forward hadron production in
$d$-$Au$ collisions at RHIC, but it fails at central rapidity
\cite{Boer:2007ug}. In particular, the logarithmic rise of $\gamma$
proved to be too slow to describe the larger $x$ central rapidity
data. Also the scaling violations of the DHJ model could not be
resolved in the data. In order to investigate to what extent the RHIC
data establish the small-$x$ properties incorporated in the DHJ model,
a new model was put forward that is not only exactly
geometrically scaling, but in addition features a stronger rise of
$\gamma$ \cite{Boer:2007ug}
\begin{equation}
 \gamma_{\rm GS}(w)=\gamma_s+(1-\gamma_s)\frac{(w^a-1)}{(w^a-1)+b}\,.
\label{gamma_GS}
\end{equation}
Here, $a=2.82$ and $b=168$ were fitted to the data. This GS model
turned out to be able to describe the $d$-$Au$ data well at all
rapidities. It must be emphasized though that when restricted to the
forward data, or equivalently the smaller $x$ data, both models
describe the data equally well. From the comparison of the model
predictions to future LHC $p\,$-$Pb$ and $p\,$-$p$ data one should be
able to learn which rise is more appropriate at small $x$. Given this
uncertainty, here we will use both models to study the transverse
$\Lambda$ polarization.

To shed light on the peak in the $p_t$ distribution, we will separate
$\mathcal{P}_\Lambda$ into a $p_t$-dependent and an $x_F$-dependent
part in the following way. To good approximation the integrals in
(\ref{Plam}) are dominated by a value of $x_F/x\equiv z$ that is
independent of $p_t$ and only moderately dependent on $x_F$.  Due to
the large power $c_q$, which suppresses small ratios $z$ in the
numerator, the values of $z$ effectively probed in the numerator and
the denominator are different. We will denote the value that dominates
the numerator with $z$ and the smaller one that dominates the
denominator with $z^\prime$. Of course in the kinematic limit $x_F\to
1$, both $z$ and $z^\prime$ must become equal to 1. In the following
analysis we will stay away from this limit and assume that $x_F$ stays
smaller than roughly 0.5. Ignoring the gluonic contributions, which is
a good approximation when $x_F$ is not too small, we can approximate
(\ref{Plam}) in the following way
\begin{align}
\nonumber
\mathcal{P}_\Lambda(p_t,x_F) &\approx 
\frac{\sum_q D_{\Lambda/q}(z)\, (x_F/z)\,f_{q/p}(x_F/z,\mu_f^2)\,
f_q^\Delta(z)}{\sum_q D_{\Lambda/q}(z^\prime)\,(x_F/z^\prime)\,f_{q/p}(x_F/z^\prime,\mu_f^2)}
\\
&\hspace{3cm}\times
 \frac{N_F\left(\frac{1}{z}(p_t-k^0_t),\frac{1}{x_F\,z}\,\frac{(p_t-k^0_t)^2}{s}\right) - 
        N_F\left(\frac{1}{z}(p_t+k^0_t),\frac{1}{x_F\,z}\,\frac{(p_t+ k^0_t)^2}{s}\right) }
  {N_F\left(\frac{1}{z^\prime}p_t,\frac{1}{x_F\,z^\prime}\,\frac{p_t^2}{s}\right)}\,.
\label{Plam_app}
\end{align}
Since $z$ and $z^\prime$ are considered constant, Eq.\
(\ref{Plam_app}) now depends on $p_t$ through the function $N_F$ only.
This is true assuming the factorization scale $\mu_f$ to be constant.
Below we will mostly choose $\mu_f=p_t$ though, but this will turn out
not to make much difference. 
We further note that since $k_t^0$ is only around $0.3\;{\rm GeV}$ or
smaller for all relevant values of $z$, we can expand
$N_F\left(\tfrac{1}{z}(p_t-k^0_t)\right) -
N_F\left(\tfrac{1}{z}(p_t+k^0_t)\right)$ in terms of $k_t^0/p_t$,
requiring $p_t \geq 1$ GeV throughout this paper,
\begin{equation}
  N_F\left(\tfrac{1}{z}(p_t-k^0_t)\right) -  N_F\left(\tfrac{1}{z}(p_t+k^0_t)\right)
\approx -2 \,\frac{k^0_t}{z}\,\frac{{\rm d}\, N_F}{{\rm d}\,q_t}\,.
\label{NFdiff}
\end{equation}
Here we have suppressed the explicit dependence on $x_2$ for
convenience and we will do so frequently below. 
Writing the dipole scattering amplitude in terms of a dimensionless function $\tilde{N}_F$,
\begin{align}
 N_F(q_t,x_2)&
\equiv\frac{2\pi}{q_t^2} \tilde{N}_F(w=q_t/Q_s(x_2),x_2)\,,
\label{Ntilde}
\end{align}
we can express Eq.~(\ref{NFdiff}) in the following way,
\begin{equation}
  N_F\left(\tfrac{1}{z}(p_t-k^0_t)\right) -  N_F\left(\tfrac{1}{z}(p_t+k^0_t)\right)
\approx2 \,\frac{k^0_t}{p_t} \frac{2\pi}{q_t^2}(2
\tilde{N}_F(w)-w\tilde{N}^\prime_F(w))\,.
\end{equation}
Using this result, we can split off the $p_t$-dependence
of the transverse polarization and write
\begin{equation}
\mathcal{P}_\Lambda(p_t,x_F) \approx 
\frac{\sum_q D_{\Lambda/q}(z)\, (x_F/z)\,f_{q/p}(x_F/z,\mu_f^2)\,
f_q^\Delta(z)/z}{\sum_q D_{\Lambda/q}(z^\prime)\,(x_F/z^\prime)
\,f_{q/p}(x_F/z^\prime,\mu_f^2)} \frac{k^0_t}{Q_s}
\frac{z^2}{z^{\prime\,2}}
F(w,w^\prime)\,,
\label{Plam_app2}
\end{equation}
where we have defined the $p_t$-dependent part of
$\mathcal{P}_\Lambda$ as a separate function $F(w,w^\prime)$,
\begin{equation}
F(w,w^\prime)=\frac{2}{w}
\frac
{2\tilde{N}_F(w)-w\tilde{N}^\prime_F(w)}
{  \tilde{N}_F(w^\prime)}.
\end{equation}

\begin{figure}[htb]
\centering
\includegraphics*[width=110mm]{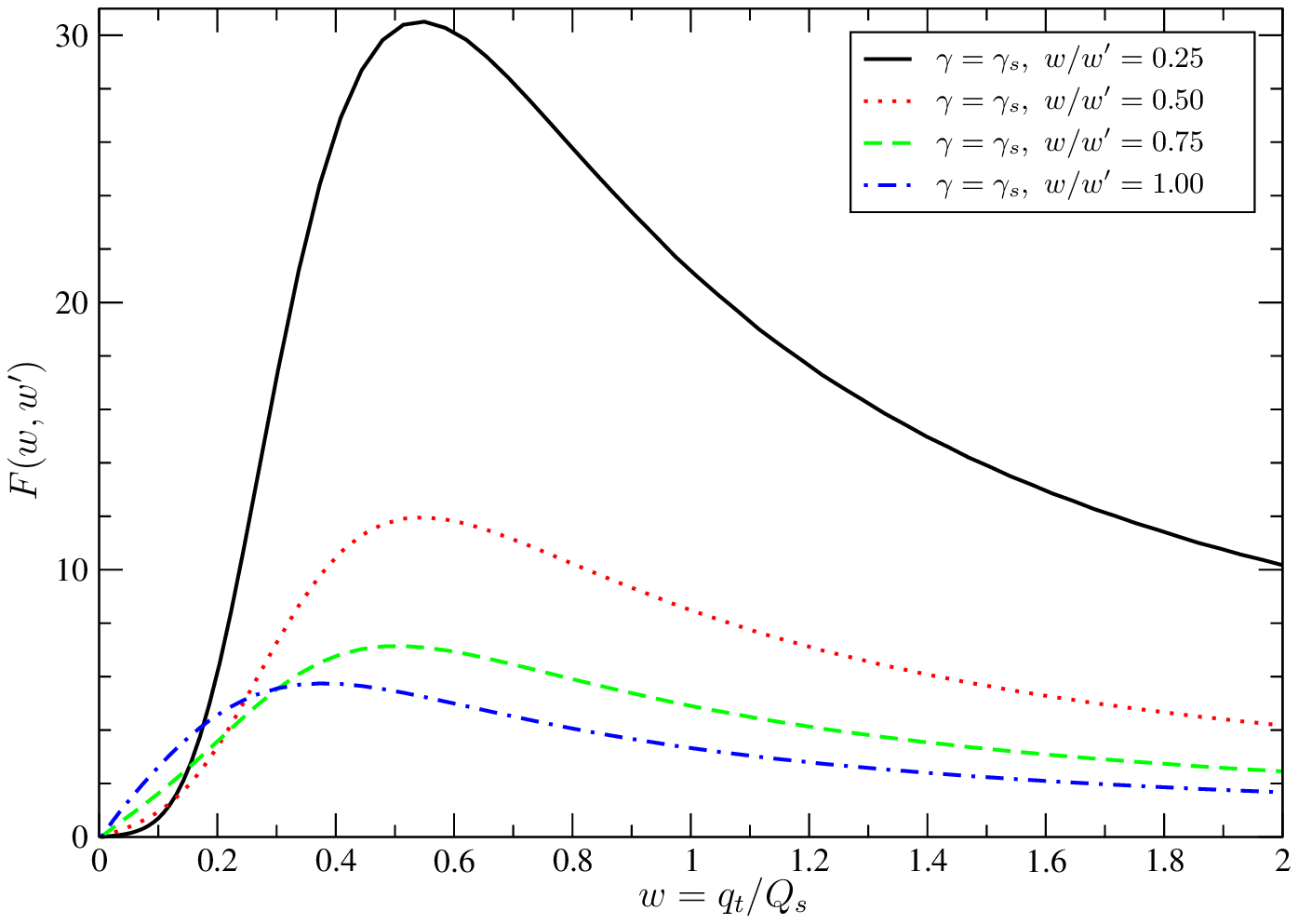}
\caption{\label{fig_Fyyp}  The function $F(w,w^\prime)$ 
for $\gamma=0.6275$ and various ratios $w/w^\prime$}.
\end{figure}

\mbox{From} the asymptotic behavior of $F$ it can be seen that it must
have an extremum. From Eqs.\ (\ref{NF_param}) and (\ref{Ntilde}), it
follows that $\tilde{N}_F\propto 1/w^{2\gamma}$ for large $w$, and
hence that $F(w,w^\prime)$ will approach $2(1+\gamma)/w$. On the other
hand, in the deep saturation regime the function (\ref{Ntilde}) is
proportional to $w^2$, so that $F(w,w^\prime)$ vanishes as $w\to0$.
Therefore, the function $F(w,w^\prime)$, and consequently $P_\Lambda$,
must have a peak as it connects these two asymptotic
behaviors. Without saturation there could also be a peak in
$P_\Lambda$, but one would in that case not expect the extremum to be
rather sharply peaked at a perturbative scale of a few GeV. Such a
peak would be a sign of saturation, especially if it increases towards
larger transverse momenta with increasing energy. Of course, there
could be a plateau-like extremum, as it appears to be the case at low
energies. However, the MV model calculation of Ref.~\cite{Boer:2002ij}
clearly shows there to be a pronounced peak, with a position
proportional to the (constant) saturation scale $Q_s$.  If this
proportionality holds when $Q_s$ evolves with $x$, the location of the
peak in $p_t$ would be a direct probe of the running of $Q_s$ through
its dependence on $x_F$. If however the peak position depends also
{\it explicitly} on $x_F$, the running of $Q_s$ cannot be
reconstructed from the peak position. Because the probed values of
$z^\prime/z=w/w^\prime$ depend on $x_F$, this means that we have to
check that they do not influence the position of the peak.
Fig.~\ref{fig_Fyyp} shows $F(w,w^\prime)$ for various values of values
of $w/w^\prime=z^\prime/z$ ranging from $0.25$ to $1$, using a dipole
scattering amplitude with a constant $\gamma=0.6275$. The curves
indeed have a clear maximum\footnote{Since the polarized part of the
fragmentation function $f_q^\Delta$ is negative for the $u$ and $d$
quarks that lead to the dominating contributions,
$\mathcal{P}_\Lambda$ will have a negative valued minimum, which for
convenience will sometimes also simply be referred to as a peak.}  as
a function of $w$. The position of the peak hardly depends on
$w^\prime/w$ if $w^\prime/w$ is not too close to 1, i.e.\ away from
the kinematic limit $x_F\to1$. Hence, we conclude that the peak of $F$
is located at an approximately constant value of $w$.  As mentioned,
this means that the minimum of $\mathcal{P}_\Lambda$ does not
explicitly depend on $x_F$, but only through the saturation scale
$Q_s$.
                                      
We find that all this remains true not only for different constant
$\gamma$'s, but also for the DHJ and GS models. The GS model actually
leads to the same peak position as constant $\gamma=\gamma_s$, because
it differs only little from $\gamma_{\rm GS}(w=1)=\gamma_s$ in the
saturation region $w\le 1$, where the peak of $F$ is located. The DHJ
model gives a slightly different peak, one that depends on the 
continuation into the saturation region $q_t<Q_s$, as will be
discussed further below.

One can estimate the $x_F$ dependence of the peak of the
resulting $p_t$-distribution as follows. Since the peak of $F$ is at a
constant value of $w=zp_t/Q_s$, where $z$ is roughly constant as well, the peak
in $p_t$ is directly proportional to $Q_s(x)$. Because the dominant
value of $x$ that is probed 
depends on both $p_t$ and $x_F$, the peak position $p_t^{\rm peak}$ will
depend on $x_F$. As the probed value of $z=x_F/x$ in the integrals in
Eq.\ (\ref{Plam}) is to good approximation constant, the target 
momentum fraction $x_2$, which sets the saturation scale $Q_s(x_2)$, 
is given by
\begin{equation}
 x_2=x\,\exp(-2\,y_h)=\frac{x}{x_F^2}\,\frac{p_t^2}{s}
\propto\frac{1}{x_F}\,\frac{p_t^2}{s}\,.
\label{x2_eq}
\end{equation}
Using this relation, we can estimate the $x_F$-dependence of the peak
position $p_t^{\rm peak}$ of $\mathcal{P}_\Lambda$. Assuming that the
saturation scale is given by a power law in $1/x$, Eq.\ (\ref{Qsx2}),
we see that
\begin{align}
 p_t^{\rm peak}&\propto Q_s(x_F,p_t^{\rm peak})
\propto Q_0\,\lf(\frac{x_F\,x_0\,s}{(p_t^{\rm
       peak})^2}\right)^{\lambda/2}\ 
\\
\Rightarrow \ p_t^{\rm peak}(x_F) &\propto Q_0x_F^{\,\lambda^\prime/2}  \,
\lf(\frac{x_0\,s}{Q_0^2}\right)^{\lambda^\prime/2}\,,
\quad\lambda^\prime=\frac{\lambda}{1+\lambda}
\label{pt-x_F}
\end{align} 
Hence, we conclude that the running of the peak position with
$x_F$ is a clear indication of the running of the saturation scale
$Q_s(x_2)$. Moreover, the power $\lambda$ can be reconstructed from
the behavior of the peak position as a function of $x_F$.

The parameterization of $Q_s$ in Eq.\ (\ref{Qsx2}) is not just based
on the GBW model fit to DIS data. The specific power law dependence on
$1/x$ stems from theoretical arguments. Small-$x$ evolution equations,
such as the GLR \cite{GLR}, BFKL \cite{BFKL} and BK \cite{BK}
equations in the fixed coupling constant case, result in such a
behavior and determine the power $\lambda$ (typically they yield
$\lambda \approx 0.9$). In the running coupling case the functional
form of $Q_s$ is different. But over the limited range of
experimentally accessible values of $x$, $Q_s$ can still be
approximated by a power law like behavior. In this way the specific
value of $\lambda=0.3$ that best describes the DIS data, can be
effectively accomodated. This implies that
$\lambda$ may be different in $p\,$-$A$ collision where a different
kinematic range is probed.

In the discussion above we have expanded $N_F$ in terms of $k_t^0/p_t$
requiring $p_t \geq 1$ GeV, so that $p_t$ is in the perturbative
regime. This should be considered a minimal requirement for the
present dipole description to be applicable and also from the
perspective of the scale choice $\mu_f=p_t$ it is a sensible lower
bound. Therefore, below we will only discuss results for which
$p_t^{\rm peak}$ is in the perturbative regime.

\section{Transverse $\Lambda$ polarization results}
\label{sect_results}
Here we will present our numerical estimates of the transverse
polarization (\ref{Plam}). For the fragmentation functions we have
chosen the leading order (LO) functions given in
\cite{deFlorian:1997zj} and for the
parton distributions the CTEQ5 LO ones \cite{Lai:1999wy}. In
\cite{DHJ2} it was shown that certain effects of higher order can be
taken into account by the DGLAP evolution of fragmentation functions
and parton distributions at the scale set by $p_t$. Unless stated
otherwise, we will therefore set the factorization scale to
$\mu_f=p_t$. We will return to the $\mu_f$ dependence of the results
later on. 

\begin{figure}[htb]
\centering
\includegraphics*[width=110mm]{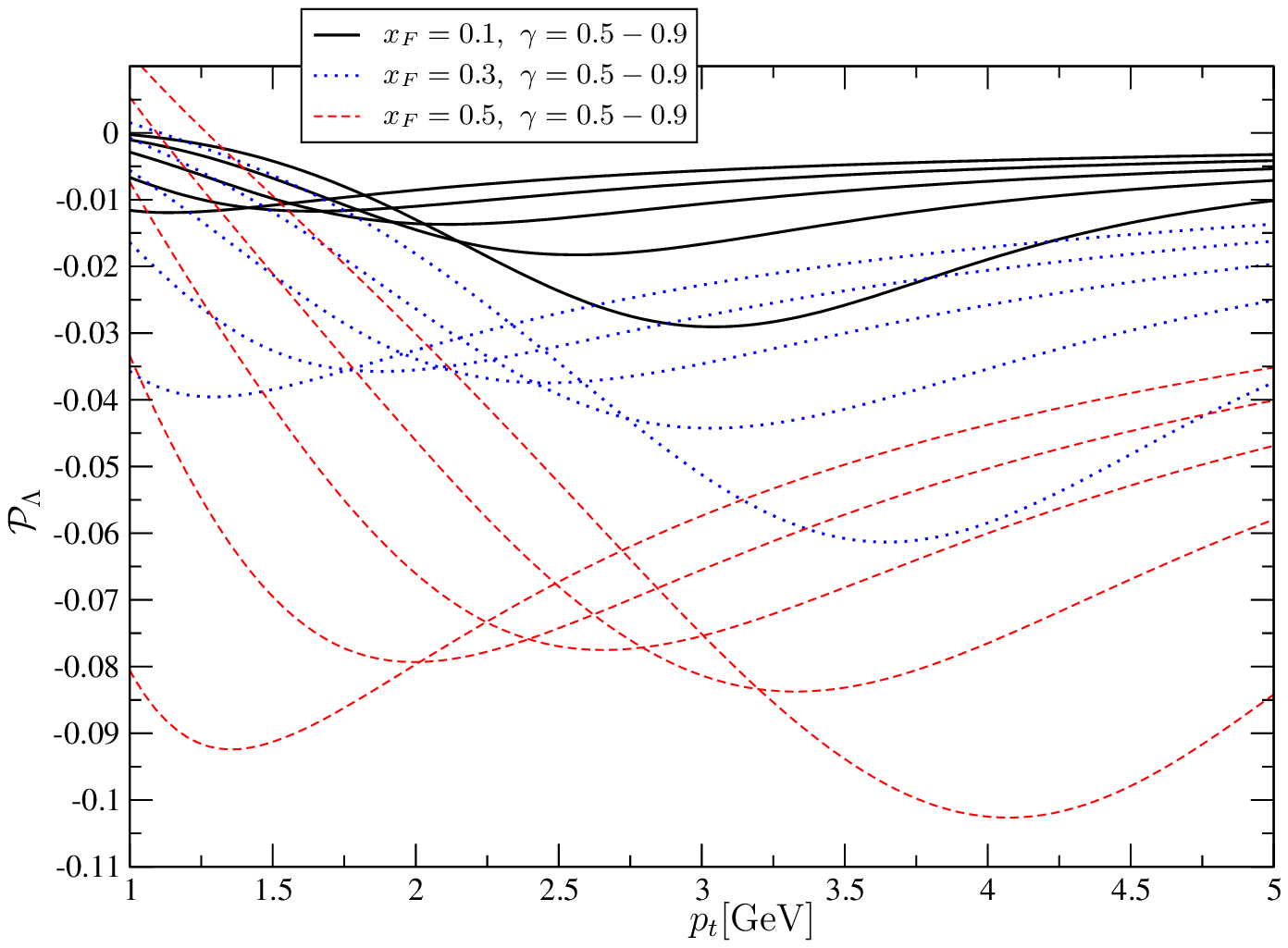}
\caption{\label{fig_Plam_g0i} $\mathcal{P}_\Lambda$ for various
  constant $\gamma$ for $p\,$-$Pb$ collisions at $\sqrt{s}=8.8\;{\rm
  TeV}$.  Curves for smaller $\gamma$ have their minimum at
  smaller $p_t$.}
\end{figure}

We first discuss the $p_t$ distribution of $\mathcal{P}_\Lambda$ for
constant values of $x_F$.
Here we will first give the result for $p\,$-$Pb$ collisions at LHC at
$\sqrt{s}=8.8\;{\rm TeV}$ explicitly and later point out how they
compare to $p\,$-$p$ collisions at LHC and $d$-$Au$ collisions at
RHIC. For the saturation scale we will use the GBW parameterization
(\ref{Qsx2}) with $Q_0^2=2.7\;{\rm GeV}^2$ instead of $1\;{\rm GeV}^2$
by taking $A_{\rm eff}=20$.  Figure~\ref{fig_Plam_g0i} shows the
resulting $\mathcal{P}_\Lambda$. It has been calculated for dipole
scattering amplitudes with various constant values of $\gamma$ from
0.5 to 0.9. The increasing magnitude of the polarization with
increasing $x_F$ is due to the polarized part of the fragmentation
$f_q^{\Delta}$ (for larger $x_F$ larger values of $z$ are probed).
The anticipated rise of the peak position with $x_F$ can be clearly
observed. Furthermore, the peak position rises approximately linearly
with $\gamma$ for all considered values of $x_F$.

\begin{figure}[htb]
\centering
\includegraphics*[width=110mm]{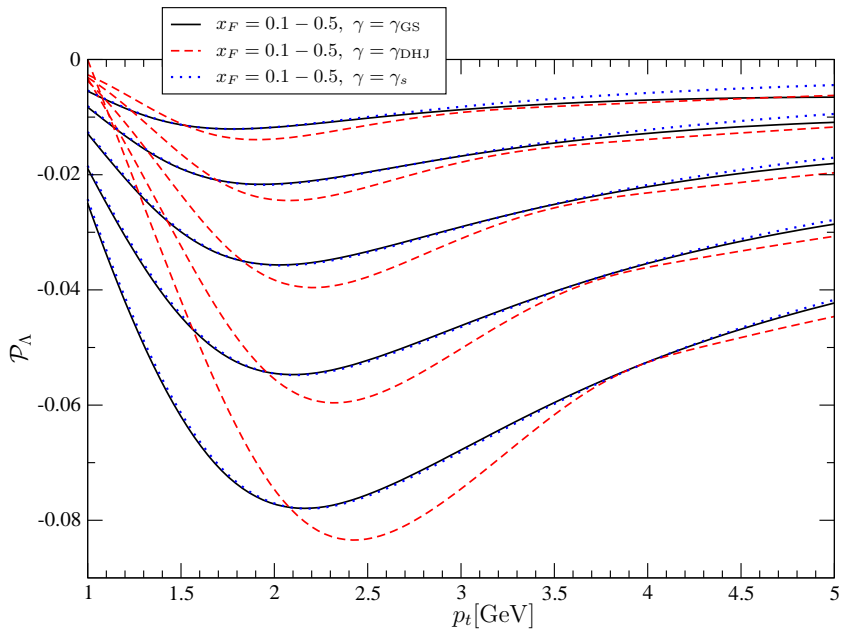}
\caption{\label{fig_Plam_alDHJgs} $\mathcal{P}_\Lambda$ 
in $p\,$-$Pb$ collisions at $\sqrt{s}=8.8$ TeV,
for  $\gamma_{\rm DHJ}$, $\gamma_{\rm GS}$ and a constant 
$\gamma_s$. The top lines
  correspond to $x_F=0.1$, the lowest to $x_F=0.5$.}
\end{figure}

Figure~\ref{fig_Plam_alDHJgs} shows the polarization for various $x_F$
as a function of $p_t$, but now for three $\gamma$'s that are all
equal at the saturation scale: a constant $\gamma_s$, $\gamma_{\rm
DHJ}$ and $\gamma_{\rm GS}$. As expected, the difference between the
polarization for $\gamma_s$ and $\gamma_{\rm GS}$ is very small
because in the saturation region $\gamma_{\rm GS}$ differs only mildly
from the value $\gamma_{\rm GS}(q_t=Q_s)=\gamma_s$. Because of the
uncertainty in the continuation of the DHJ parameterization in the
saturation region, the estimate of $\mathcal{P}_\Lambda$ around the
peak is ambiguous. If we would for instance continue $\gamma_{\rm
DHJ}$ by keeping it constant for $q_t<Q_s$, we would obtain roughly
the same result as for $\gamma_{\rm GS}$. The fact that the DHJ and GS
models yield similar results for the observed behavior of the peak
indicates our findings are rather robust and to a certain extent model
independent. In contrast, the magnitude of the polarization is subject
to considerable uncertainty, mostly due to the parameterization of
$\Delta^N D$, but also somewhat due to the choice of the
factorization scale.

\begin{figure}[htb]
\centering
\includegraphics*[width=110mm]{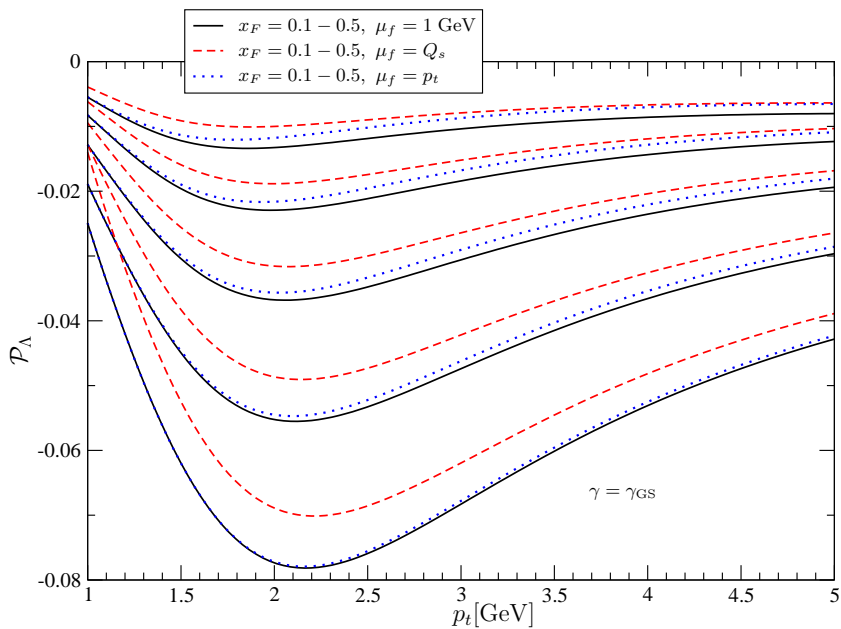}
\caption{\label{fig_Plam_al_frac} $\mathcal{P}_\Lambda$ in
  $p\,$-$Pb$ collisions at LHC for the
  scaling $\gamma_{\rm GS}$ (\ref{gamma_GS}) and three different
  choices of the factorization scale $\mu_f=p_t, Q_s$ and 1 GeV.  The top lines
  correspond to $x_F=0.1$, the lowest to $x_F=0.5$.}
\end{figure}

Thus far we have used a factorization scale $\mu_f=p_t$.  However, in
Ref.\ \cite{Boer:2002ij} $\mu_f=Q_s$ was considered, which may also be
a natural choice. In the present case that would lead to an
$x$-dependent factorization scale. Figure~\ref{fig_Plam_al_frac} shows
$\mathcal{P}_\Lambda$ with $\gamma_{\rm GS}$ for three different
factorization scales, $\mu_f=p_t$, $\mu_f=Q_s$, and a constant scale
$\mu_f=1$ GeV.  As can be seen, for constant $x_F$ the shape of the
$p_t$ distributions is rather independent of the factorization scale.
The normalization does depend on the choice of $\mu_f$, but
still only moderately. Therefore, choosing $\mu_f=p_t$ does not
noticeably affect our claim that the $x_F$ dependence of the peak
momentum directly probes the $x$ dependence of the saturation scale.

\begin{figure}[htb]
\centering
\includegraphics*[width=110mm]{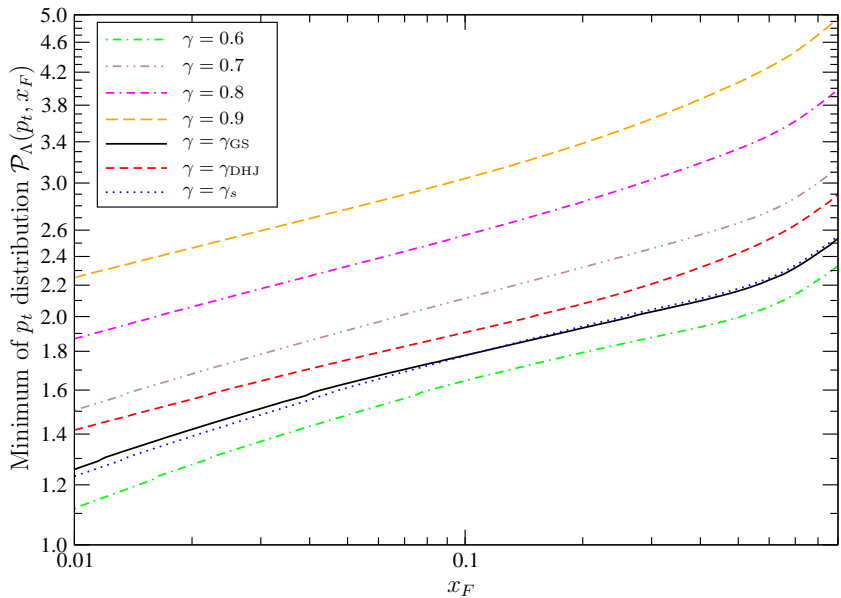}
\caption{\label{fig_xf_mupt_ptmin} The peak position of the $p_t$
  distribution $\mathcal{P}_\Lambda(p_t,x_F)$ as a function of $x_F$
  in double logarithmic representation for various choices of $\gamma$.}
\end{figure}

Fig.~\ref{fig_xf_mupt_ptmin} shows the $x_F$ dependence of the peak
position of the $p_t$ distribution for various choices of $\gamma$.
The lines for constant $\gamma$ confirm that the peak position scales
linearly with $\gamma$. Moreover, for not too large $x_F$, the power
law rise of $p_t^{\rm peak}$ with $x_F$ is consistent with the result
we obtained in Eq.\ (\ref{pt-x_F}), including the fact that the power
is independent of $\gamma$. As expected, the results for $\gamma_{\rm
  GS}(w)$ (\ref{gamma_GS}) and the constant $\gamma=\gamma_s$ are very
close to each other. The curve for $\gamma_{\rm DHJ}$ (\ref{gammaDHJ})
is similar to that of a constant $\gamma$ that is slightly larger than
$\gamma_s$. This is because $\gamma_{\rm DHJ}$ rises
rather quickly in the saturation region as $q_t$ decreases.

As can be seen from Fig.~\ref{fig_xf_mupt_ptmin} all slopes are also
numerically consistent with the power $\lambda'/2$ in Eq.\
(\ref{pt-x_F}) for $\lambda=0.3$. This implies that an increase in
$x_F$ by a factor 5 leads to a shift in the peak position of
approximately 20\%. This can be seen directly in
Fig.~\ref{fig_Plam_al_frac} too, when comparing the peak position at
$x_F=0.1$ and 0.5. Unfortunately, this is not a large shift, but it
does give an estimate for the precision with which the peak position
needs to be determined. It should be mentioned that, as discussed
before, the value of $\lambda$ may be different in $p\,$-$A$
collisions than the one taken from the analysis of the small-$x$ DIS
data. A larger value of $\lambda$ would of course result in a stronger
$x$ dependence of $Q_s$ and therefore in a larger $x_F$ dependence of
the peak position that would be easier to observe. At small $x_F$,
where the position of the peak is less pronounced, it will be harder
to determine than at large $x_F$. The value of $p_t/Q_s$ at which the
peak is situated depends --too good approximation linearly-- on
$\gamma$ in the saturation region $q_t\le Q_s$. As can be seen in
Fig.~\ref{fig_Plam_alDHJgs}, the peak is located for $\gamma_{\rm GS}$
and $\gamma=\gamma_s$ at almost the same postion. We find empirically
that in these cases the minimum in the $p_t$ distribution shows up at
$w \approx 0.55$, i.e.\ $p_t\approx 0.55\,z\,Q_s$, where $z$ rises
slightly with $x_F$ from 0.7 to 1 in the limit $x_F\to 1$. Depending
on the continuation of the DHJ model to the saturation region the peak
is situated at a different $w$. For the continuation (\ref{gammaDHJ})
the peak shows up at $p_t\approx 0.60\,z\,Q_s$, since it rises again
towards smaller $q_t$ in the saturation region.
   
Similar results for $\mathcal{P}_\Lambda$ are obtained for $p\,$-$p$
collisions at LHC and $d\,$-$Au$ collisions at RHIC. Again an extremum
is observed at around one half times the saturation scale that shows
the same $x_F$ dependence as for $p\,$-$Pb$ scattering at
LHC. However, due to the different kinematics and targets, $Q_s$ and
hence the position of the peak $p_t^{\rm peak}$ is in both cases
lower. Following the same line of arguments leading to the $x_F$
dependence of $p_t^{\rm peak}$ (\ref{pt-x_F}), one can estimate its
$\sqrt{s}$ and $Q_0$ dependence,
\begin{equation}
 p_t^{\rm peak}(x_F,\sqrt{s^\prime},Q_0^\prime)=p_t^{\rm peak}(x_F,\sqrt{s},Q_0)
\lf(\frac{Q_0^\prime\,\sqrt{s^\prime}^\lambda}{Q_0\,\sqrt{s}^\lambda}\rg)^{1/(1+\lambda)}\,.
\end{equation}
For $p\,$-$p$ at LHC $Q_0$ is 1 GeV and $\sqrt{s}=14\;{\rm
TeV}$. Hence, the peak position is expected to be reduced by a factor
of 1.3 with respect to $p\,$-$Pb$ collisions at $\sqrt{s}=8.8\;{\rm
TeV}$. An explicit calculation confirms that this estimate works very
well, i.e.\ the $x_F$-dependent extremum is expected to show up 
approximately between 1.5 and 2.0 GeV for $x_F \sim
0.1-0.5$. For $d\,$-$Au$ collisions at RHIC the probed values
of $x_2$ are less small due to the smaller energy. Hence, the probed
values of $Q_s$ and hence $p^{\rm peak}$ are reduced even more, namely
by a factor of 2.4 compared with $p\,$-$Pb$ collisions at LHC, which
may situate it below the perturbative regime $p_t\gtrsim 1\;{\rm
GeV}$, even for constant values of $x_F=0.1-0.5$. However, given the
uncertainties in e.g.\ the values of $Q_0$ and $\lambda$, a peak in
the perturbative region is not ruled out, especially for larger
$x_F$. From this perspective it may still be worthwhile to investigate
this observable at RHIC.

Up to now we focused on the calculation of $\mathcal{P}_\Lambda$ at
constant $x_F$, where the dependence on $\sqrt{s}$ is not that
large. However, from an experimental point of view it might be more
convenient to measure $\mathcal{P}_\Lambda$ for constant rapidities
$y_h$. As demonstrated before, there is a clear peak in the $p_t$
distribution at fixed $x_F$, which is at different locations for
different $x_F$. Therefore, since at fixed $y_h$ a range of values of
$x_F$ contributes, the peak will be smeared out to some extent (this
can also be observed for the DHJ model predictions of single spin
asymmetries in forward pion production in the collisions of
transversely polarized protons with unpolarized protons
\cite{Boer:2006rj}).  Hence, it is not clear {\em a priori\/} whether
the peak remains observable and whether the peak position is still a
clear probe of the saturation scale.

\begin{figure}[htb]
\centering
\includegraphics*[width=110mm]{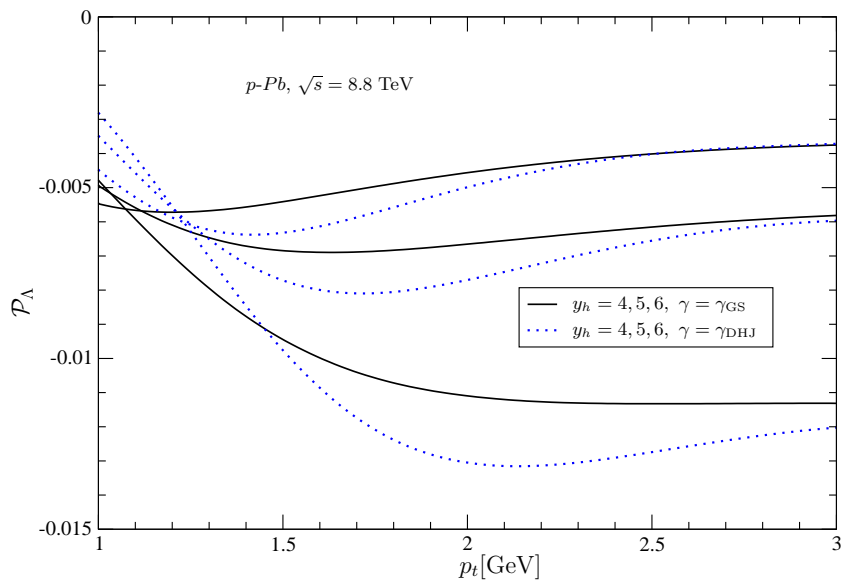}
\caption{\label{fig_Plam_LHC_alDHJ} $\mathcal{P}_\Lambda$ in
  $p\,$-$Pb$ collisions at $\sqrt{s}=8.8$ TeV, for constant $y_h$
  using  $\gamma_{\rm GS}$ and $\gamma_{\rm DHJ}$. The top lines
  correspond to $y_h=4$, the lowest to $y_h=6$.}
\end{figure}

For LHC kinematics, we know from the previous analysis that a peak at
transverse momenta larger than 1 GeV requires
$x_F=p_t/\sqrt{s}\,\exp[y_h]\gtrsim0.01$. At LHC such a peak is thus
only expected in the forward region $y_h\gtrsim 4$.
Fig.~\ref{fig_Plam_LHC_alDHJ} shows $\mathcal{P}_\Lambda$ for
$p\,$-$Pb$ scattering at LHC at $\sqrt{s_{NN}}=8.8$ TeV, for values of
$y_h=4,5,6$. Indeed, the extremum is in these cases located at a
$p_t$ larger than 1 GeV, but it is much less pronounced than at fixed
$x_F$ and for the GS model less recognizable than for the DHJ
model. We also note that the magnitude of the asymmetry is 
considerably reduced compared to the fixed $x_F$ case.

At RHIC the saturation scale becomes roughly of the order $Q_s
\simorder 1$ GeV for forward $\Lambda$'s with rapidities of around 4.
However, unlike for the MV model, the peak position for the DHJ and GS
models is located considerably below $Q_s$, that is, below $p_t = 1$
GeV. An explicit calculation of $P_\Lambda$ for RHIC confirms that
even for $y_h=4$ a peak is not expected to be above 1 GeV. In other
words, $Q_s(x)$ can presumably not be extracted in a trustworthy
manner from a fixed $y_h=4$ analysis at RHIC, unless $Q_0$ and/or
$\lambda$ turn out to be larger than expected at present.

\section{Conclusions}
The transverse polarization of $\Lambda$ particles displays a peak at
the saturation scale when described using the MV model for the dipole
scattering amplitude. We find that in the more realistic
case where the dipole amplitude depends on $x$, such a peak in the
$p_t$ distribution remains. The position of the peak, $p_t^{\rm
peak}$, is still proportional to $Q_s$, and therefore offers a
direct experimental probe of this scale. For fixed values of $x_F$,
the $x$ dependence of $Q_s$ can be reconstructed from the $x_F$
dependence of $p_t^{\rm peak}$. It would be very interesting to
compare the function $Q_s(x)$ obtained in $p\,$-$A$ collisions in this
way with the GBW model one that was obtained from DIS data, in
order to establish consistency among the descriptions of all available
data. The power $\lambda$ in $Q_s \sim x^{-\lambda/2}$ determines how
strongly the peak varies with $x_F$. Using $\lambda=0.3$ as obtained
from DIS, which according to a dipole scattering description is
compatible with forward hadron production $d\,$-$Au$ data of
RHIC, we have obtained the following results. In $p\,$-$Pb$ collisions
at LHC, for values of $x_F$ that are between 0.1 and 0.5, the position
of the peak is expected between $p_t=1.5$ and 2.5 GeV. This
result is obtained for a range of dipole models that
includes the DHJ and GS models. In $p\,$-$p$ collisions, the position
of the peak is reduced by a factor of 1.3, but is still in the
perturbative regime. In $d\,$-$Au$ collisions at RHIC, the position
of the peak is smaller by a factor of 2.4 with respect to $p\,$-$Pb$
at LHC, due to the much smaller energy. Hence, observing the peak in
the perturbative regime at RHIC seems unlikely, except perhaps at even
larger $x_F$ values. 

For fixed values of the rapidity instead of $x_F$, the peak in the
$p_t$ distribution gets smeared out and is reduced in size. Moreover,
in this case the polarization peaks in the perturbative regime
$p_t\gtrsim1$ GeV only for $\Lambda$ rapidities of 4 or larger in
$p\,$-$Pb$ collisions at LHC. Therefore, $\Lambda$ polarization
LHC data at fixed $x_F$ are best suited for the purpose of
establishing the $x$-dependence of $Q_s$ in $p\,$-$A$ collisions.

Even though the presented quantitative estimates are to some
extent model dependent, the qualitative features of the $\Lambda$
polarization, i.e.\ the position of the peak with respect to $Q_s$ and
its running with $x_F$, are expected to be generic for the small-$x$
region. This offers a unique possibility to probe $Q_s$ directly in
$p\,$-$A$ collisions.

\begin{acknowledgments}
We thank Les Bland for fruitful discussions.
This research is part of the research program of the ``Stichting voor
Fundamenteel Onderzoek der Materie (FOM)'', which is financially supported
by the ``Nederlandse Organisatie voor Wetenschappelijk Onderzoek (NWO)''.
\end{acknowledgments}

\end{document}